\documentclass[manuscript]{aastex}

\usepackage{graphicx}   
\usepackage{amsmath}    
\usepackage{xcolor}     

\shorttitle{AutoClassMK}
\shortauthors{Short}

\begin{document}


\title{AutoClassMK:  A public neural network for automatic 2D MK classification of normal stars in basic Python}


\author{C. Ian Short}
\affil{Department of Astronomy \& Physics and Institute for Computational Astrophysics, Saint Mary's University,
    Halifax, NS, Canada, B3H 3C3}
\email{ian.short@smu.ca}





\begin{abstract}

We present {\tt AutoClassMK}, a simple, fully-connected, five-layer 
double-headed neural network
written entirely in Python and Numpy that classifies {\it normal} stellar spectra
conforming to the {\tt libr18} MK atlas of \citet{libr18} in the 2$D$ MK
classification system with a
high degree of precision and recall.  
        {\tt AutoClassMK} has the distinction
of having transparent basic code with no calls to specialized libraries.
In this paper we take care to explicitly describe in detail the ideas and operations
that enable the network.
	Training AutoClassMK required
us to develop large, noisy artificial training and test sets by
augmenting the {\tt libr18} and {\tt libr18\_27} MK atlases and to simplify the luminosity
classification so that every combination of spectral- and luminosity-class
is represented in the training set.  We then test the network's ability to
predict the MK spectral type of noisy augmentations of spectra in the
	{\tt libr18\_225} MK atlas.
We then implemented the same architecture in
PyTorch to gain further insight and to enable execution on CUDA GPU's.
All codes and the training and test sets are available from the OpenStars www site:
www.ap.smu.ca/OpenStars.
 
\end{abstract}



\section{Introduction}

A star's Morgan-Keenan (MK) spectral type is ideally an independent parameter that is independently determined so that any correlations
with other observable quantities or fitted modeling quantities can be subject to on-going reinvestigation.
Therefore, the classification is ideally strictly empirical and is based entirely on a program spectrum that conforms with
the MK standard:  A wavelength range ($\Delta\lambda$) of $\sim$3900 to $\sim$4900 $\AA~$, a spectral resolving power ($R$) 
of $\sim$2000, a reciprocal linear dispersion (${{\Delta\lambda}\over{\Delta x}}$) of $\sim67 \AA~$ mm$^{-1}$ (and, originally,
a glass plate detector with the Kodak OJ III photographic emulsion and a spectral image with a widening of $\sim$1 mm.)
The classification criterion is strictly morphological:  Overall qualitative visual pattern matching against an MK atlas such 
as that of \citet{abt68}, or, currently, a digital MK atlas such as {\tt libr18} \citep{libr18}.
Therefore, automation of MK classification can benefit from the application of neural networks (NN's), which can model an arbitrarily complex function
relating sample content to probabilities corresponding to sample class.
For ''normal'' stars, the
basic system is 2$D$ and the complete MK spectral type consists of the spectral class (SC, including numerical subclass) and the luminosity class (LC).

\paragraph{}

The purpose of {\tt AutoClassMK} is to de-mystify and publicize the basic methodology of simple NN's,
especially back-propagation, given their increasing importance
to astronomy yet their opaque nature.  A distinguishing property of {\tt AutoClassMK}
is that it has been developed entirely in basic Python and Numpy and every step is transparent for inspection and modification
and the code could be a suitable starting point for a student project.
To that end, {\tt AutoClassMK} is available on the OpenStars www site (www.ap.smu.ca/OpenStars). 

\paragraph{}

NN's that automatically classify stellar spectra have already been developed dating back to the pioneering work of 
\citet{weaver} and \citet{bailer} 
who developed and investigated simple NN's like the one we present here.  More recently, more sophisticated NN's incorporating 
convolution (CNN's) and principal component analysis (PCA) among other ideas that can work with large objective prism
survey data and distinguish peculiar stars, among other things, have been presented (see, for example, \citet{han}, \citet{mca} (MCA-Net), 
\cite{SFNet} (SFNet), and \citet{wu}).  

\paragraph{}

In Section \ref{tsets} we describe our large augmented training and test sets, in Section \ref{network} we describe 
the NN architecture and model, in Section \ref{results} we present statistics describing how well the 
NN predicts the test set during training and an independent set of samples after the NN has been trained, 
and in Section \ref{torch} we describe a PyTorch implementation.

\section{Training and test sets}
\label{tsets}

We produce an adequate training set by augmenting the libr18 atlas. 
We begin by clipping the red end of the spectra to eliminate edge effects that might 
complicate the training.  We then simplify the luminosity classification by re-labelling
LC II stars as LC III and LC IV stars as LC V.   LC Ia and Ib stars are re-labeled as
LC I.  This ensures that every combination of SC and LC is represented by at least one sample
so that there are no empty cells in the 2$D$ training set that might undermine the training.

\paragraph{}

For each spectrum in the libr18 atlas ({\it ie. primary}) spectrum, we produce 300 noisy, 
randomly erroneously calibrated variations corresponding to 300 different {\it normal} stars of identical
MK type.
\begin{enumerate}
\item{We assume a Gaussian distribution of random zeroth- and first-order horizontal registration errors corresponding 
to random small variations in zero-point positioning and linear dispersion with values of $\sigma$ of 
$0.030$ and $2.0\times 10^{-5}$ equivalent \AA, respectively.  We are effectively  
encoding wavelength, $\lambda$, with pixel number, so we do not explicitly account for errors in 
$\lambda$ scale calibration.}
\item{We assume a Gaussian distribution of random errors in the Doppler-shift
	to the lab frame with a $\sigma$ value of 1.0 km s$^{-1}$, which is additive with the horizontal registration variation.}
\item{We assume a Gaussian distribution of random errors in the zeroth-, first-, and second-order 
	polynomial fitting coefficients used for continuum rectification, with $\sigma$ values of 
	$2\times 10^{-3}$, $5\times 10^{-3}$, and $3\times 10^{-3}$, respectively, which gives rise
	to vertical variation due to a distribution of overall normalization levels and residual slopes and curvatures.}
\item{We assume a Gaussian distribution with a $\sigma$ value of 0.3 pixels for variation among the standard deviation of
	Gaussian convolution kernels used to
	broaden the primary spectrum to account for both a random distribution of macroturbulent 
	velocity dispersion, $\xi$, and rotational $v\sin i$ values.  This will disproportionately 
	affect weaker spectral features.  We take a Gaussian to be a good
	approximation for the rotational broadening kernel for slowly rotating stars.}
\item{We assume that the primary spectra are brightness-limited and their noise is that of Poisson statistics.
	We assume a gain of 5000 counts and use the Numpy routine {\tt poisson} to generate 300 noise 
	spectra that we add to the primary spectrum, increasing the vertical variation.  This is equivalent 
	to assuming a signal-to-noise of
	$\sim$70, which is a significant underestimate.   However, we deliberately overestimate the
	noise so as to train the NN to recognize noisy spectra.}
\end{enumerate}

\begin{figure}
\includegraphics[width=\columnwidth]{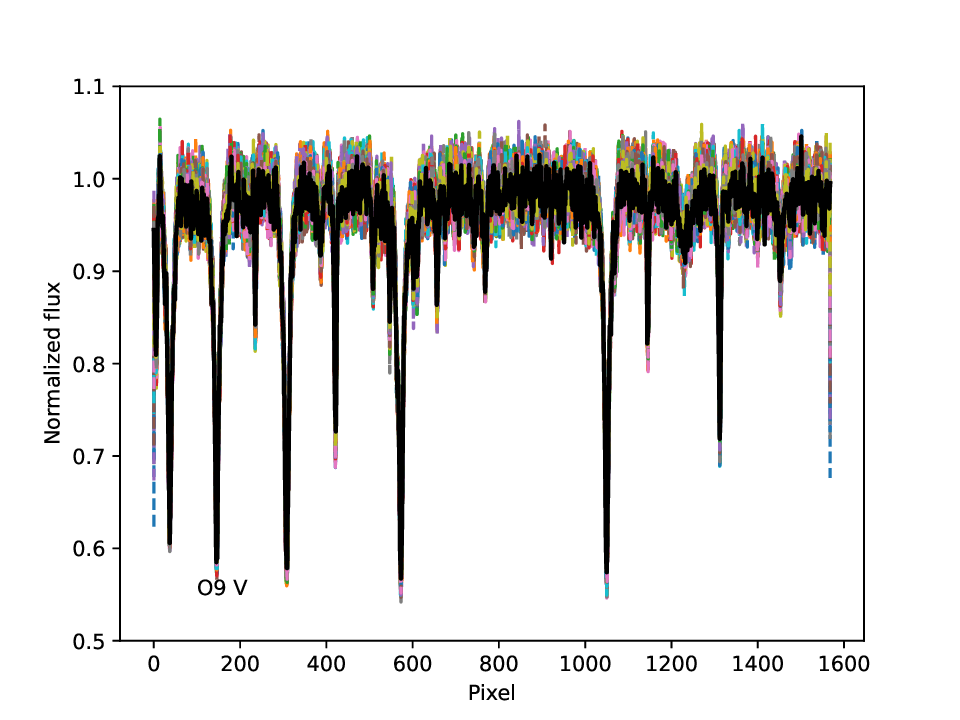}
	\caption{Original spectrum (thick black line) of class O9 V from the 
	libr18 atlas \citep{libr18} and 300 noisy, randomly mis-calibrated 
	variations (dashed colored lines).
  \label{augment}
}
\end{figure}

\begin{figure}
\includegraphics[width=\columnwidth]{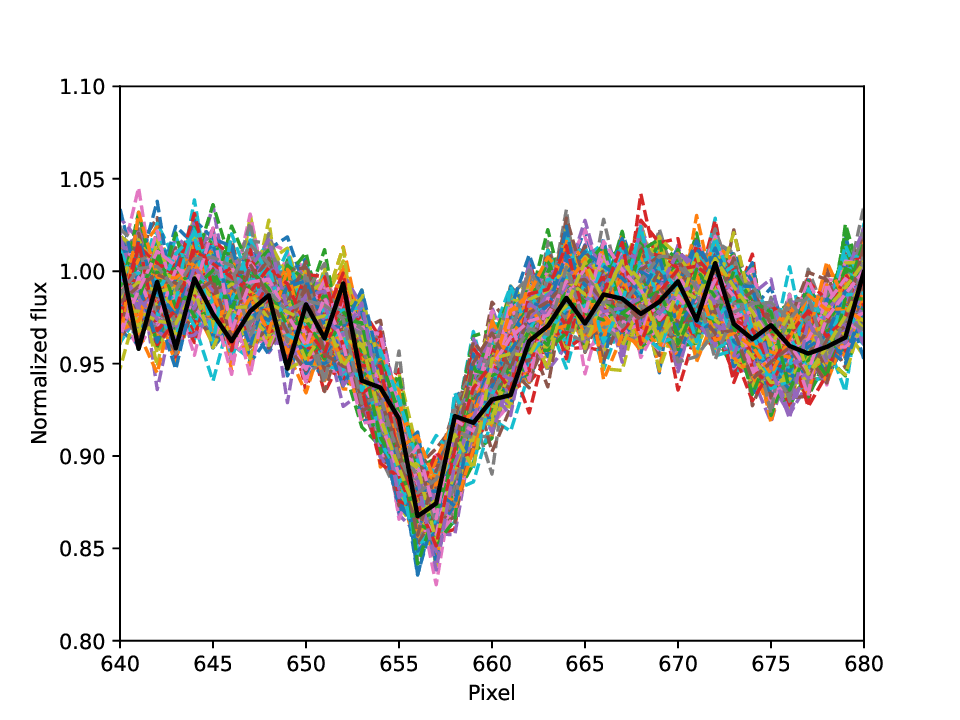}
	\caption{
		As for Fig. \ref{augment} except for a narrow region centred on pixel 660 showing the 
		effect of data augmentation on weaker spectral features.
  \label{augment2}
}
\end{figure}

We augment each of the 122 spectra in our pruned version of the libr18 atlas to produce a training set of
36600 spectra.  These are then stored in random order so that a sequence of records corresponds to a 
random selection of spectral types.  Figs. \ref{augment} and \ref{augment2} show the original spectrum and the 300
augmented spectra for MK type O9 V on the broad-band and narrow-band scale, respectively.

\paragraph{}

To assess the state of training (convergence) of the NN after each iteration, we require a {\it test} set of 
comparable samples that are independent of the training set.  We generate a test set by following the data augmentation procedure
described above to generate 30 variations of each of the 97 spectra in the libr18\_27 atlas \citep{libr18}, giving
us a test set of 2910 spectra.

\section{The network}
\label{network}

We have developed the simplest type of NN:  A {\it fully connected} (FC, or {\it dense-layer}) network (a {\it perceptron}) in which the value of each 
element in a 1$D$ array representing a sample at a particular layer of the network is generally determined by the value of {\it all}
the elements in an array representing the sample at the previous layer.  The equivalent description in the language of NN's 
is that each ''neuron'' in a given layer is affected by all the neurons in the previous layer.

\subsection{Architecture }

Our network has three hidden layers for a total of five main layers.  The input layer has 1600 neurons, one per input pixel
in the digitized training and test spectra.  
Each training and test sample consists of a set of normalized fluxes {\it vs} pixel number.  This is
tantamount to encoding physical real-number $\lambda$ values with whole numbers. 

\paragraph{}

The output layer is bifurcated with two parallel linear transformations from the second-to-last 
layer, so that the network is {\it double-headed} for 2$D$ classification.  The SC head has 31 neurons and the LC head has three neurons, corresponding to 
the 31 predictable spectral classes and the three predictable luminosity in our training set.  
Our network has a {\it tapered} architecture and for the three hidden layers we have
chosen 768, 256, and 64 for the numbers of neurons.

\paragraph{}

During training of the NN, a 1$D$ (vectorized) representation of a stellar spectrum (a ''sample'') enters the input layer, and the sample
moves forward through the NN by way of linear transformations that reduce the number of elements representing that sample.  When passing from 
layer $k$ with $m$ neurons to layer $k+1$ with $n$ neurons, the sample is transformed from being represented with vector $\vec{z}_k$ of length $m$ 
to vector $\vec{z}_{k+1}$ of length $n$ by way of the linear transformation

\begin{equation}
	\vec{z}_{k+1} = \bar{W}_k \cdot \vec{z}_k + \vec{b}_k 
\end{equation}

where $\bar{W}_k$ and $\vec{b}_k$ are the transformation matrix and the bias vector for a linear transformation {\it out of} layer $k$.  
With these transformations alone, the network can numerically model with a {\it linear} approximation a function that 
relates the sample contents (spectral features) to probabilities corresponding to classes (SC's and LC's).
Training the network involves iteratively converging the values of $\bar{W}_k$ and $\vec{b}_k$ for all layers, $k$.  The values
at either final layer (head), $\vec{z}_K$, are referred to as {\it logits}, where $K$ is the total number of layers.

\subsection{Activation function}

To allow the NN to model the {\it non}-linear relationship between the sample content (spectral features) and the probabilities
of each spectral type, we operate on the transformed sample at each layer of $k < K$ with a non-linear activation function,
$f_A$ so that

\begin{equation}
	\vec{a}_k = f_A(\vec{z}_k),
\end{equation}

where $\vec{a}_k$ is the {\it activated} representation of the sample at layer $k$.
We have experimented with two common activation functions, Rectified Linear Units (RELU$(x)$) and Gaussian Error Linear 
Units (GELU$(x)$), where $x$ is a pre-activated value.  We also require the derivative, ${df_A}\over{dx}$ for back-propagation.
The RELU function is

\begin{equation}
	f_A(x) = x, {\rm if}~ x > 0\\
\end{equation}
\begin{equation}
	   = 0, {\rm if}~ x < 0
\end{equation}

\begin{figure}
\includegraphics[width=\columnwidth]{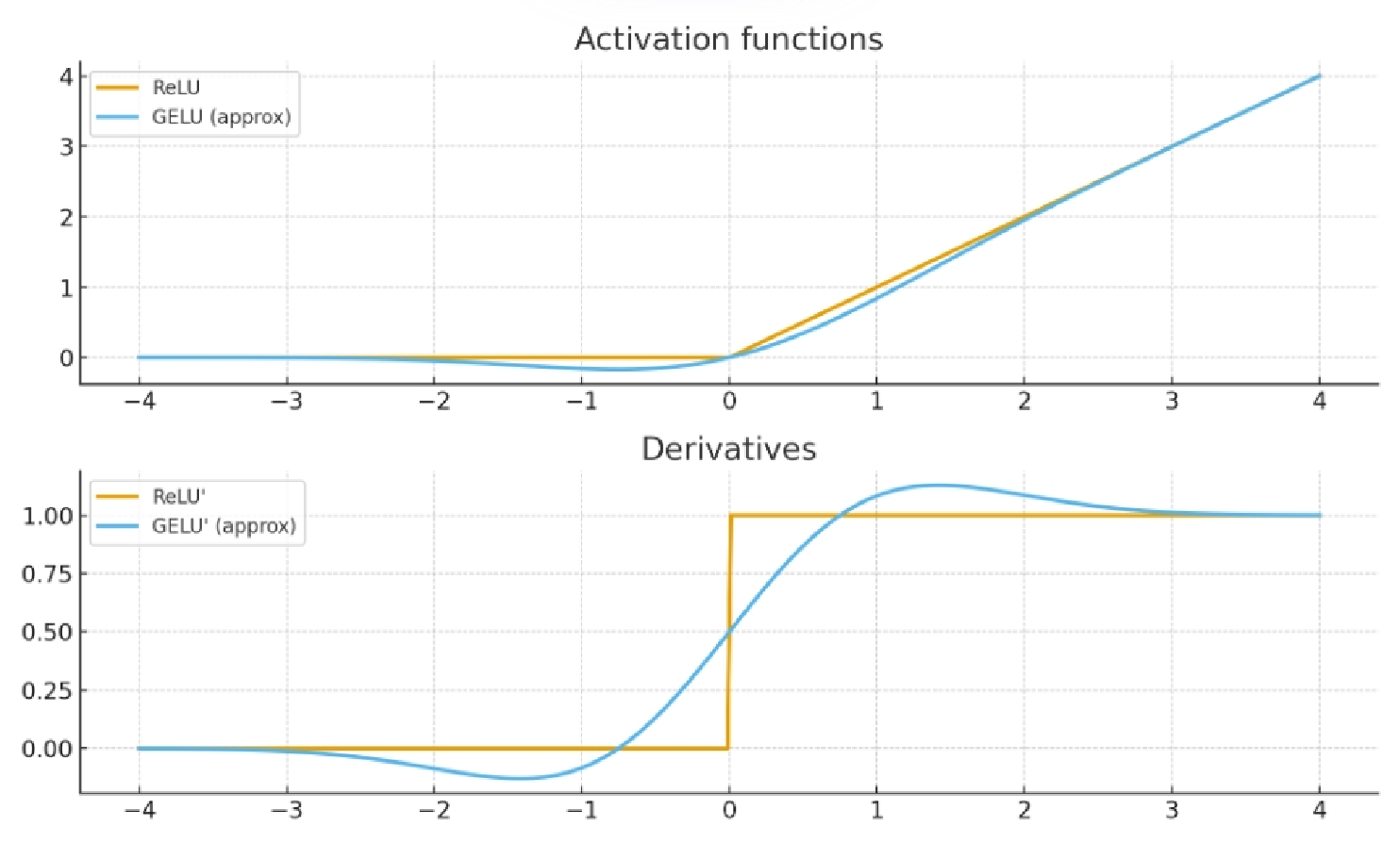}
\caption{Post- {\it vs} pre-activated element value for the Gaussian Error 
	Linear Units (GELU) 
	activation function, $f_A$, (blue) used in our NN.  Also shown for reference is
	the simpler Rectified Linear Units (RELU) activation function (orange).
	(Credit: ChatGPT (OpenAI))
  \label{fA}
}
\end{figure}

We avoid computing the Gaussian Error function by adopting the 
common approximation for the GELU function,

\begin{equation}
	f_A(x) \approx 0.5x\{1+\tanh[\sqrt{2\over\pi}(x+cx^3) ] \}
\end{equation}

where $c$ is 0.044715.  The advantage of RELU activation is that it can be computed very quickly.  However, the piecewise nature of RELU is known
to introduce unphysical jaggedness in the relationship between $\Delta\bar{W}_k$ and $\Delta\vec{z}_K$.
Both the RELU and GELU activation functions and their derivatives are shown in Fig. \ref{fA}.

\subsection{Probability distributions and loss}

For each sample, the raw logits, $\vec{z}_K$, at the final layer, $K$, at either head are normalized among themselves
by evaluating $z'_i = z_{K, i} - z_{K, {\rm max}}$ to mitigate against numerical instability.  Then, for each sample,
the vector of normalized logits, $\vec{z}_K$, is
converted into a vector of probabilities, $\vec{p}$, with the {\tt SOFTMAX} function

\begin{equation}
	\vec{p} = {\tt SOFTMAX}(z'_i) = {\exp{z'_i} \over \sum_j{\exp{z'_j}} }.
\end{equation}

The  {\tt SOFTMAX}$(z'_i)$ function is equivalent to the Boltzmann distribution for excitation equilibrium 
in the case where all the statistical weights are equal, where $z'_i$ is the negative of the excitation energy 
relative to the ground state ($z_{K, {\rm max}}$) and 
$\sum_j{\exp{z'_j}}$ is the partition function.  The {\tt SOFTMAX}$(z'_i)$ values are always positive
and will sum to unity.

\paragraph{}

The sample labels for the true SC and LC are numerically encoded as label vectors, $\vec{l}$, with the 
{\it one-hot} encoding scheme, where
a one-hot-encoded label, $\vec{l}$, is the result of operating on the true probability distribution,
$\vec{z}_{\rm true}$,
with the {\tt SOFTMAX} function, as

\begin{equation}
	\vec{l} = {\tt SOFTMAX}(\vec{z}_{\rm true})
\end{equation}

 This encodes the true SC or LC class label as a vector that is equal to zero
everywhere except equal to one in the element corresponding to the true class.
For example, in our simplified luminosity classification, if the true LC is I then the
corresponding one-hot-encoded label is the vector $\vec{l} = [ 1, 0, 0 ]$ 
{\it ie.} the sample is an LC I star with probability one and an LC III or V star with probability zero.

\paragraph{}

We implicitly evaluate the average prediction error for each batch of samples after each 
training iteration with
the cross-entropy (CE) loss, $L$, where, for a batch of $N$ samples,

\begin{equation}
	L = -\sum^{N}{\vec{l}\cdot\ln\vec{p} }
\label{L}
\end{equation}

The advantage of one-hot encoding is that in a vectorized language like Numpy, the terms in Eq. \ref{L} can
be calculated simply with a vector operator. The CE loss, $L$, is defined so that each sample's contribution to its batch's $L$ value at that iteration step
depends on the probability distribution, $\vec{p}$, and, thus, the confidence of the current prediction, as well as the
accuracy of the prediction.  One can show that the definition of the {\tt SOFTMAX} and $L$ functions is such 
that

\begin{equation}
	{{dL}\over{d\vec{z}}} = \vec{p} - \vec{l}
\label{dL}
\end{equation}

where the gradient ${{dL}\over{\vec{dz}}}$ is needed for back-propagation and can be evaluated simply.
The value, $L$, is not used explicitly in back-propagation, but its definition is implicit
because we use Eq. \ref{dL}.  We compute $L$ for reporting purposes.

\subsection{Forward pass and back-propagation}

\paragraph{Initialization } The matrices, $\bar{W}_k$, are initialized with a Gaussian distribution of random values centered on zero with
a $\sigma$ value of 0.1 for all layers, $k$.  The bias vectors, $\vec{b}_k$ are initialized to $\vec{0}$
for all $k$.

\paragraph{}

For each layer, $k$, we implement the transformation 

\begin{equation}
	\vec{z}_{k+1} = \bar{W}_k\cdot\vec{z}_k + \vec{b}_k 
\end{equation}

with Numpy's vectorized dot-product operator.  If $k$ is equal to 1 then the operand, $\vec{z}_1$, is the 
input sample, and if $k$ is equal to $K$ then the output, $\vec{z}_K$, is the final vector of raw logits
at either of the two output heads.  For the all $k$ values of $k < K$, we activate the 
current representation of the sample by implementing

\begin{equation}
	\vec{a}_k = f_A(\vec{z}_k)
\end{equation}

For the layer $k=K-1$ we implement two parallel transformations for the SC and LC heads.

\subsubsection{Back-propagation}

The batch-averaged error for the current predictions is

\begin{equation}
\label{errzK}
	\vec{\delta z}_K = {{\vec{p} - \vec{l}}\over N_{\rm samples}}
\end{equation}

Then, the corresponding errors in the matrix $\bar{W}_K$ and the vector $\vec{b_K}$ are 

\begin{equation}
\label{errWK}
    \bar{\delta W}_K = \vec{a}^T_{K-1}\cdot\vec{\delta z}_K
\end{equation}

\begin{equation}
\label{errbK}
	\vec{\delta b}_K = \sum^{N_{\rm samples}}  \vec{\delta z}_K
\end{equation}

Eqs. \ref{errzK} through \ref{errbK} are executed twice, once for each of the SC and LC heads. The errors
from the two heads are combined as

\begin{equation}
	\vec{\delta a}_{K-1} = \vec{\delta z}_{K, \rm SC}\cdot\bar{W}^T_{K, \rm SC} + \vec{\delta z}_{K, \rm LC}\cdot\bar{W}^T_{K, \rm LC}  
\end{equation}

Subsequently, for all layers going backward through the network to decreasing $k$ we have

\begin{equation}
	\vec{\delta a}_{k-1} = \vec{\delta z}_k\cdot\bar{W}^T_k
\end{equation}

\begin{equation}
	\vec{\delta z}_k = \vec{\delta a}_k\cdot {{df_A(\vec{z}_k)}\over {dz}}
\end{equation}

\begin{equation}
\label{errW}
    \bar{\delta W}_k = \vec{a}^T_{k-1}\cdot\vec{\delta z}_k
\end{equation}

\begin{equation}
\label{errb}
        \vec{\delta b}_k = \sum^{N_{\rm samples}} \vec{\delta z}_k
\end{equation}

For the special case of $k=2$ we have

\begin{equation}
    \bar{\delta W}_2 = \vec{z}^T_1\cdot\vec{\delta z}_2
\end{equation}

where $\vec{z}_1$ is the input sample.

\subsection{Optimization step and learning rate}

For each batch of samples for which we calculate batch-average $\bar{\delta W}_k$ and $\vec{\delta b}_k$
values, we perform an $n^{\rm th}$ update, or optimization step, for all layers, $k$, as 

\begin{equation}
	\bar{W}^{n+1}_k = \bar{W}^n_k - \lambda\bar{\delta W}_k
\end{equation}

and

\begin{equation}
	\vec{b}^{n+1}_k = \vec{b}^n_k - \lambda\vec{\delta b}_k
\end{equation}

where $\lambda$ is the {\it learning rate} and serves as a damping parameter to mitigate against
oscillations that might prevent convergence or lead to divergence.   Typical values of $\lambda$
for 1$D$ FC NN's range form 0.01 to 0.001.  One can make use of a {\it learning rate schedule}
to smooth convergence, where a typical schedule is one in which the value of $\lambda$ is
decreased by a factor of 2 every 50 outer iterations.

\subsection{Mini-batch stochastic gradient descent (SGD)}

The values of $\bar{\delta W}_k$ and $\vec{\delta b}_k$ used in the optimization step are averages computed
for mini-batches of $N_{\rm samples}$ randomly drawn samples and reflect the average current prediction error for a random, approximately representative
subset of the overall training set.  If $M$ is the total number of samples in the training set, then the number of mini-batches,
$N_{\rm batch}$, is $M/N_{\rm samples}$.  Iterative convergence is performed with a doubly nested loop in which the inner loop 
over mini-batches is executed $N_{\rm batch}$ times while the values of the $\bar{W}_k$ and $\vec{b}_k$ are 
cumulatively refined.  The mini-batch loop is embedded in an outer loop over {\it epochs}, where each epoch 
the samples are randomly reshuffled among the mini-batches.  If there are $N_{\rm epoch}$ epochs
then the values of the $\bar{W}_k$'s and $\vec{b}_k$'s are cumulatively refined $N_{\rm epoch}\times N_{\rm batch}$
times, learning from $N_{\rm epoch}\times N_{\rm batch}$ unique randomly drawn mini-batches.  Training with mini-batches
rather than the overall training set greatly increases the number of optimization steps that occur for a 
given amount of processing.

\paragraph{}

The value of $N_{\rm samples}$ is chosen to be just small enough that a given mini-batch
has a distribution of spectral types that is only approximately representative of the overall training set.
This allows the iterative convergence to proceed stochastically, and provides noisy dithering that mitigates against 
the solution becoming trapped in a local shallow minimum. 

\section{Results}
\label{results}

\begin{figure}
\includegraphics[width=\columnwidth]{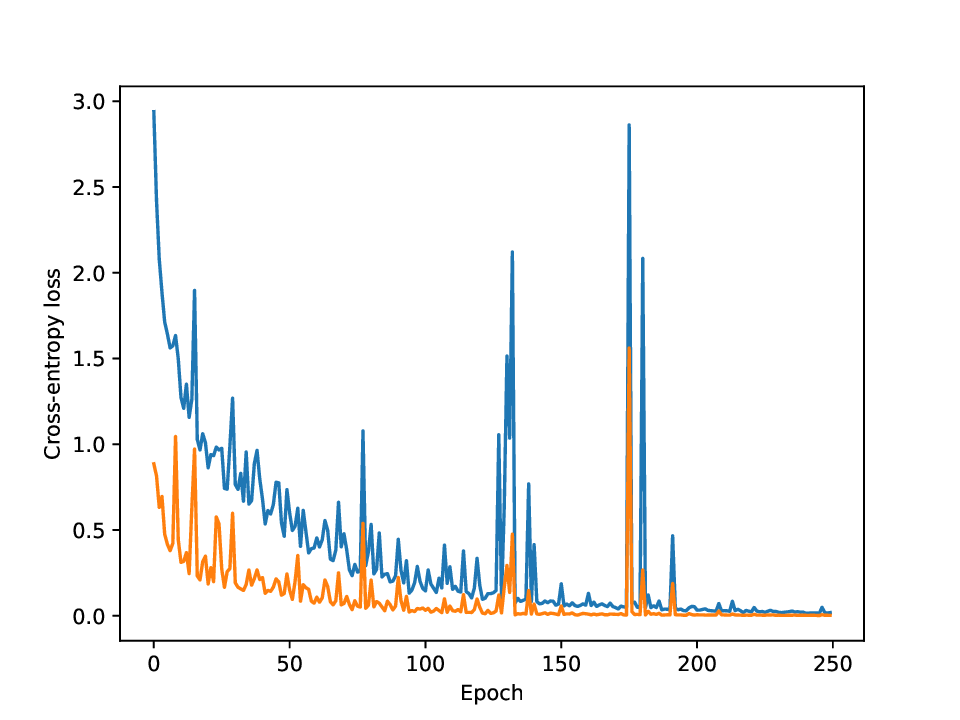}
	\caption{Mini-batch cross-entropy (CE) loss, $L$, for the last mini-batch {\it vs} outer
	iteration for our NN training with 250 outer iterations and
	a mini-batch size of 200 samples for the SC (blue) and
	LC (orange) classification.
  \label{loss}
}
\end{figure}

\begin{figure}
\includegraphics[width=\columnwidth]{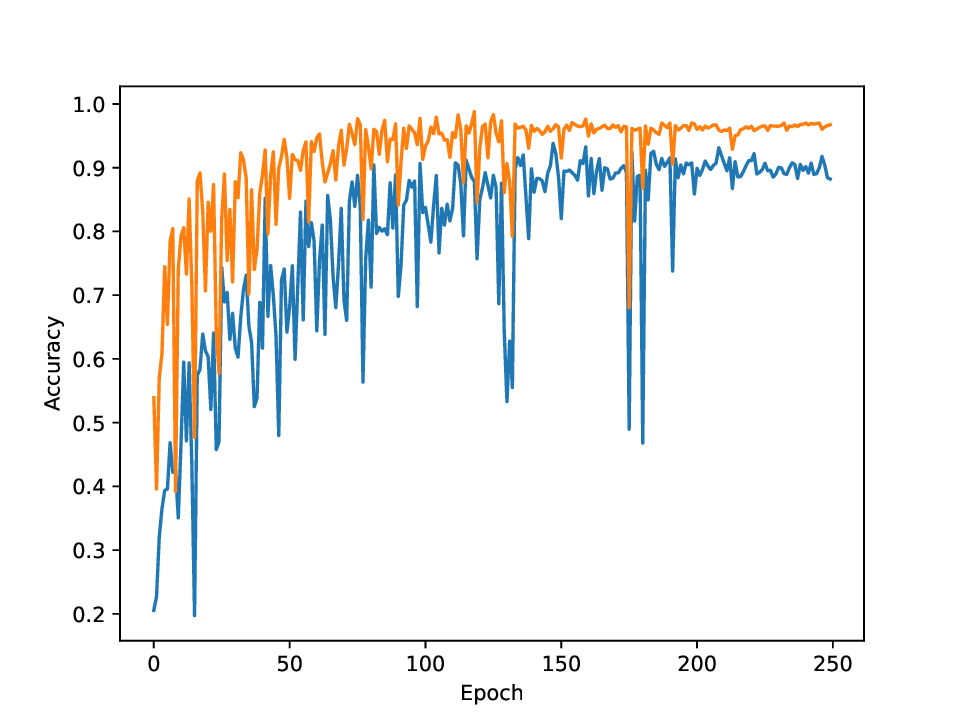}
	\caption{Same as Fig. \ref{loss} except that the dependent
	quantity is test-set prediction accuracy ({\tt acc}).
  \label{acc}
}
\end{figure}

Figs. \ref{loss} and \ref{acc} show the CE loss, $L$, and the accuracy ({\tt acc}), respectively, {\it vs} iteration number for our
SC and LC training.  After a relatively rapid early decrease in $L$, and corresponding increase in {\tt acc}, during the first 50 - 100 
iterations, the rate of convergence slows 
and asymptotically approaches limiting values of $L$ of $\sim 0.018$ for SC and $\sim 0.002$ for LC, and limiting values of {\tt acc} of 
$\sim 0.88$ for SC and $\sim 0.97$ for LC, within $\sim$100 iterations.  The convergence 
is non-monotonic on the scale
of a few iterations, and our experience is that a learning-rate schedule in which the value of $\lambda$ is
decreased by a factor of 2 every 50 outer iterations can smooth the convergence, but at a cost of significantly increasing 
the number of iterations required.

\paragraph{}

In what follows $N_{\rm TP}$, $N_{\rm FP}$, and $N_{\rm FN}$ denote the numbers of true positive (TP), false positive (FP),
and false negative (FN) classifications. 
Table \ref{stat} shows the {\tt acc} value, the {\it precision} ($N_{\rm TP}/(N_{\rm TP}+N_{\rm FP}$)), the {\it recall} 
($N_{\rm TP}/(N_{\rm TP}+N_{\rm FN}$)), and the $F1$ score ($2\times$precision$\times$recall/(precision+recall))
computed with the {\tt sklearn.metrics.classification\_report} module
for the SC and LC, evaluated using our {\it test} set of 2910 augmented noisy samples from the libr18\_27 atlas.  
Consistent with the behavior seen in Figs. \ref{loss}
and \ref{acc}, we see that the average $F1$ score is greater for the LC (0.97) than it is for the SC (0.86).

\begin{table}[h!]
  \begin{center}
    \caption{Training statistics.}
    \label{stat}
    \begin{tabular}{r|r|r|r}

\multicolumn{4}{l}{Spectral Classification (SC)}\\  
	    & Precision & Recall    & F1 \\
acc         &           &           & 0.88\\
Average     & 0.89      & 0.87      & 0.86\\
\multicolumn{4}{l}{Luminosity Classification (LC)}\\  
	    & Precision & Recall    & F1 \\
acc         &           &           & 0.97 \\
Average     & 0.97      &  0.87     & 0.97 \\

    \end{tabular}
  \end{center}
\end{table}

\subsection{Prediction}

\begin{figure}
\includegraphics[width=\columnwidth]{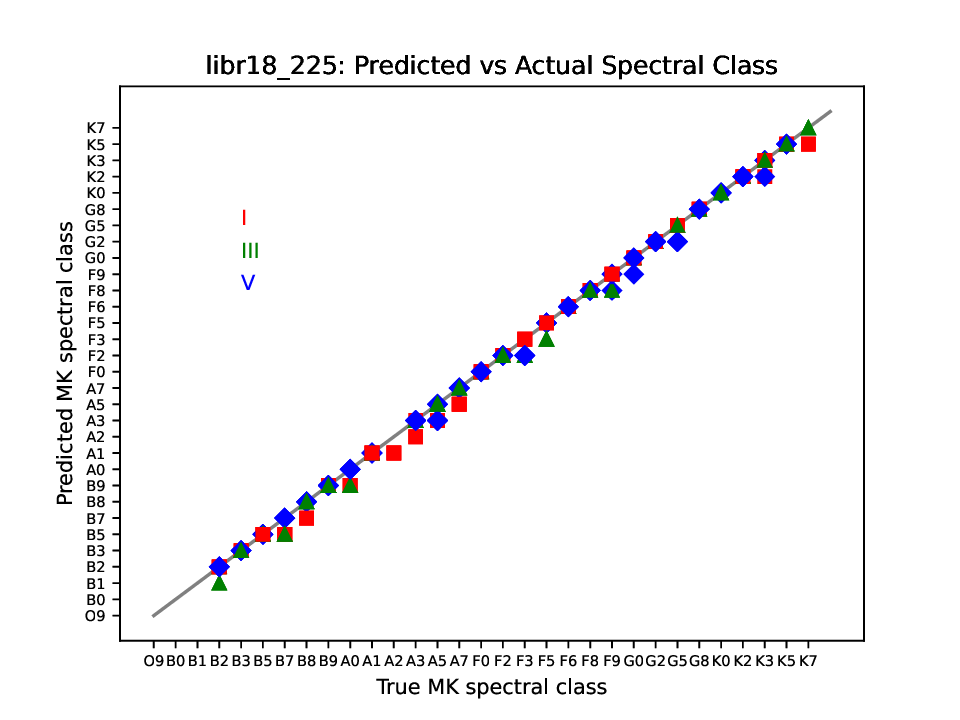}
	\caption{Predicted {\it vs} true SC for noisy augmentations of the libr18\_225
	\citep{libr18} atlas for luminosity class V (blue diamonds), III (green triangles), and I (red squares) stars.  
	The gray line indicates the locus of prefect prediction.
  \label{predSC}
}
\end{figure}

Querying the trained NN (prediction) required executing the forward pass stage of the training described in Section
\ref{network}.  Samples must conform to the parameters of the {\tt libr18} atlas in that the spectra must be {\it registered} such that 
the $\lambda$-range 
3800-4600 \AA~ spans 1600 pixels and samples must be continuum-rectified to unit continuum.
We test our trained NN by using it to predict the SC and LC of augmented spectra taken from the libr18\_225 atlas \citep{libr18}.
We follow the same augmentation procedure of Section \ref{tsets} to produce three noisy, randomly mis-calibrated
variations for each of the 85 spectra in our pruned version of libr18\_225, for a total of 255 spectra in the set 
that our NN attempts to classify.  Fig. \ref{predSC} shows the predicted SC {\it vs} the true SC for each of the three
LC values. 
Generally, the predicted SC is within one represented sub-class of the true SC.

\paragraph{} 

\begin{figure}
\includegraphics[width=\columnwidth]{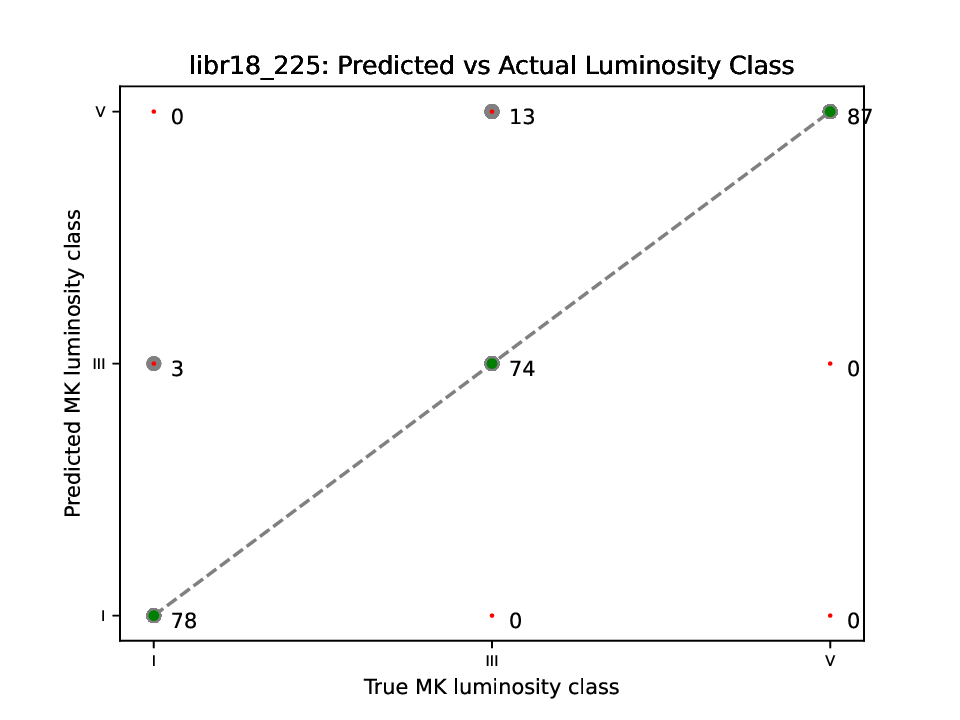}
	\caption{The $3\times 3$ confusion matrix for predicted {\it vs} true
	LC for noisy augmentations of the libr18\_225
        \citep{libr18} atlas.  The locus of perfect prediction is the anti-diagonal.
	The annotations are the numbers of spectra with each combination of 
	true and predicted LC.
  \label{predLC}
}
\end{figure}

 Fig. \ref{predLC} shows a $3\times 3$ confusion matrix for the LC.
We can see that the LC has excellent precision and recall for LC V and I.  All 87 of the spectra with a true LC of V 
had correct LC predictions, and of the 81 spectra of true LC I, all but three had correct LC predictions.
Our NN had greater difficulty distinguishing LC III stars from LC V stars - of the 87 spectra of true
LC III, 13 of them were predicted to be of LC V.

\section{PyTorch implementation}
\label{torch}

To gain further insight into our NN and to facilitate execution on GPU's equipped with CUDA, 
we implemented the same architecture in the PyTorch deep learning library \citep{pytorch}.  The PyTorch implementation of our five-layer 
FC NN is shown in Table \ref{PyT1}.

\begin{table}[h!]
  \begin{center}
	  \caption{{\tt AutoClassMK} PyTorch architecture.}
    \label{PyT1}
    \begin{tabular}{rcl}

	& $z_{\rm SC}, z_{\rm LC} = {\tt\rm MODEL(z_1)}$ & \\
	    CE$_{\rm SC}={\tt\rm CELoss}(z_{\rm SC}, z^{\rm True}_{\rm SC})$ & & CE$_{\rm LC}={\tt\rm CELoss}(z_{\rm LC}, z^{\rm True}_{\rm LC})$ \\
	    &  ${\rm CE}_{\rm TOTAL} = {(w_{\rm SC}{\rm CE}_{\rm SC}+w_{\rm LC}{\rm CE}_{\rm LC})\over (w_{\rm SC}+w_{\rm LC})}$  & \\
            & {\tt SGD.ZERO\_GRAD(MODEL, $\lambda$)} & \\
	    & {\tt CE}$_{\rm TOTAL}${\tt .BACKWARDS()} & \\
	& {\tt SGD.STEP(MODEL, $\lambda$)} & \\

    \end{tabular}
  \end{center}
\end{table}

where $w_{\rm SC}$ and $w_{\rm LC}$ are parameters that allow us to weight the relative importance of the errors
of the SC and LC predictions in the training.  Given that the NN has greater difficulty learning the SC- than the 
LC-dimension, we have found it advantageous to adopt $w_{\rm SC}=2$ and $w_{\rm LC}=1$ to force the NN to 
learn more from the SC prediction error during training. 
The {\tt MODEL} is shown in Table \ref{PyT2}.

\begin{table}[h!]
  \begin{center}
	  \caption{{\tt AutoClassMK} PyTorch model ({\tt MODEL} of Table \ref{PyT1}).}
    \label{PyT2}
    \begin{tabular}{rcl}

	    &       $z_1$ = {\tt LINEAR}(1600, 768) & \\
	    &       $a_1$ = {\tt GELU}($z_1$)   & \\
	    &       $z_2$ = {\tt LINEAR}(768, 256)  &\\
	    &       $a_2$ = {\tt GELU}($z_2$)   & \\
	    &       $z_3$ = {\tt LINEAR}(256, 64)  &\\
	    &       $a_3$ = {\tt GELU}($z_3$)   & \\
    $z_{\rm SC} = ${\tt LINEAR}(64, 31)  &  & $z_{\rm LC} = ${\tt LINEAR}(64, 3)\\
    $z_{\rm SC} = ${\tt SOFTMAX}($z_{\rm SC}$) & & $z_{\rm LC} = ${\tt SOFTMAX}($z_{\rm LC}$) \\

    \end{tabular}
  \end{center}
\end{table}

The PyTorch version is implemented in a {\it device agnostic} way and will automatically run on a GPU
if it detects a CUDA processor on the host.  We have successfully run the PyTorch version on both a CPU and a GPU of
the Digital Research Alliance of Canada (DRAC) machine nibi. 

\section{Future Work}

{\tt AutoClassMK} and the accompanying training and test sets provide ample opportunities for experiments that a student could 
do with nothing other than the basic Python and Numpy installations:

\begin{itemize}
	\item{Studying how training convergence and the precision and recall of the trained network depend on
		the noisiness, mis-calibration variance, and number of samples per class, of the training set and
		on the noisiness of the set for which the classes are being predicted.}
	\item{Studying the how training convergence depends on arbitrary parameters such as the numbers of neurons in the hidden layers,
		the learning rate, $\lambda$, and the size of the mini-batches, $N_{\rm samples}$.}
	\item{More substantially, studying the effect of adding and removing hidden layers.}
\end{itemize}

\acknowledgements{}

We used ChatGPT (OpenAI) for conceptual discussion and code prototyping. All final implementations and scientific interpretations were performed by the author.
Fig. \ref{fA} was produced by ChatGPT (Open AI).
This research was enabled in part by support provided by ACENET and the Digital Research Alliance of Canada (DRAC, alliancecan.ca).
Some data preprocessing and numerical calculations were performed using the NumPy library, available at https://numpy.org, in Python.


\begin{thebibliography}{00}

\bibitem[Abt {\it et al.}(1968)]{abt68} Abt, H. A, Meinel, A. B., Morgan, W. W., Tapscott, I. W., 1968, {\it An Atlas of Low Dispersion Grating Spectra}, Tuscon 
\bibitem[Bailer-Jones, Irwin \& von Hippel(1998)]{bailer} Bailer-Jones, C.A.L., Irwin, M. \& von Hippel, T., 1998, \mnras, 298, 361
\bibitem[Fu {\it et al.}(2024)]{SFNet} Fu, H., Liu, P., Qi, X., Mei, X., 2024, Research in Astronomy and Astrophysics, 24, 095023
\bibitem[Gray \& Corbally(2014)]{libr18} Gray, R. O. \& Corbally, C. J., 2014, \aj, 147, 80
\bibitem[Han, Kang \& Jung(2026)]{han} Han, S., Kang, W. \& Jung, J-H, 2026, Astronomy and Computing, 54, 101024
\bibitem[Li (2025)]{mca} Li, H., 2025, Experimental Astronomy, 60, 17
\bibitem[Paszke {\it et al.}(2019)]{pytorch} Paszke, A., {\it et al.}, 2019, arXiv:1912.01703v1 (PyTorch)
\bibitem[Weaver \& Torres-Dodgen(1997)]{weaver} Weaver, Wm. B., Torres-Dodgen, A. V., 1997, \apj, 487, 847 
\bibitem[Wu {\it et al.}(2024)]{wu} Wu, J., He, Y., Wang, W., Qu, M., Jiang, B., Zhang, Y., 2024, \aj, 167, 260
\end{thebibliography}
\end{document}